\def\mytitle{Multisection in the Stochastic Block Model using Semidefinite Programming}

\def\myauthors{Naman Agarwal\footnote{namana@cs.princeton.edu,
Computer Science, Princeton University}, Afonso S.
Bandeira\footnote{bandeira@mit.edu, Department of Mathematics, Massachusetts Institute of Technology (most of the work presented in this paper was conducted while this author was at Princeton University). ASB acknowledges support from AFOSR Grant No. FA9550-12-1-0317}, Konstantinos
Koiliaris\footnote{koiliar2@illinois.edu, Computer Science, University of
Illinois, Urbana - Champaign}, Alexandra Kolla\footnote{akolla@illinois.edu,
Computer Science, University of Illinois, Urbana - Champaign}}
\def\myappendix{Appendix}
\documentclass[11pt]{article}
\usepackage[fullpage,nousetoc,hylinks]{paper}
\theoremstyle{plain}

\BEGINDOC
\begin{titlepage}
\makeheader
\begin{abstract}
We consider the problem of identifying underlying community-like structures in
graphs. Towards this end we study the Stochastic Block Model (SBM) on
$k$-clusters: a random model on $n=km$ vertices, partitioned in $k$ equal
sized clusters, with edges sampled independently across clusters with
probability $q$ and within clusters with probability $p$, $p>q$. The goal is to
recover the initial ``hidden'' partition of $[n]$. We study semidefinite
programming (SDP) based algorithms in this context. In the regime $p =
\frac{\alpha \log(m)}{m}$ and $q = \frac{\beta \log(m)}{m}$ we show that a certain natural SDP based
algorithm solves the problem of {\em exact recovery} in the $k$-community SBM,
with high probability, whenever $\sqrt{\alpha} - \sqrt{\beta} > \sqrt{1}$, as
long as $k=o(\log n)$. This threshold is known to be the information
theoretically optimal. We also study the case when $k=\theta(\log(n))$. In
this case however we achieve recovery guarantees that no longer match the
optimal condition $\sqrt{\alpha} - \sqrt{\beta} > \sqrt{1}$, thus leaving
achieving optimality for this range an open question. 
\end{abstract}
\end{titlepage}
\pagebreak
%
\section{Introduction}

Identifying underlying structure in graphs is a primitive question for scientists: can existing communities be located in a large graph?
Is it possible to partition the vertices of a graph into strongly connected
clusters? Several of these questions have been shown to be hard to answer, even
approximately, so instead of looking for worst-case guarantees attention
has shifted towards average-case analyses. In order to study such questions, the
usual approach is to consider a random \cite{Mc01} or a semi-random \cite{FK01,MMV14} generative model of 
graphs, and use it as a benchmark to test existing algorithms or to develop new ones. With respect to identifying underlying community
  structure, the Stochastic Block Model (SBM) (or planted partition model)
  has, in recent times, been one of the most popular choices. Its growing popularity is largely due to the 
  fact that its structure is simple to describe, but at the same time it has interesting
  and involved phase transition properties which have only recently been
  discovered
  (\cite{DKMZ11,MNS12,MNS13,ABH14,CX14,MNS14,HWX14,HWX15,
  AS15,Bandeira_Laplacian_2015}).
  
In this paper we consider the SBM on $k$-communities defined as follows. Let $n$ be a multiple of $m$,
$V=[n]$ be the set of vertices and $P=\{P_i\}$ be a partition of them into
$k$ equal sized clusters each of size $m = \frac{n}{k}$. Construct a random
graph $G$ on $V$ by adding an edge for any two vertices in the same cluster independently with probability $p$ and any
two vertices across distinct clusters independently with probability $q$ where
$p>q$.
We will write $G \sim \mathcal{G}_{p,q,k}$ to denote that a graph $G$ is
generated from the above model. Given such a $G$ the goal is
to recover (with high probability) the initial hidden partition $P$.

The SBM can be seen as an extension of the Erd\H{o}s-R\'enyi random graph model
\cite{ER} with the additional property of possessing a non-trivial underlying
community structure (something which the Erd\H{o}s-R\'enyi model lacks). This
richer structure not only makes this model interesting to study theoretically, but also renders it closer to real world inputs, which tend to have a community structure.
It is also worth noting that, as pointed out in \cite{CX14}, a slight generalization of the SBM encompasses several classical planted random graph problems including planted clique
\cite{AKS98}, \cite{Mc01}, planted coloring \cite{AK97}, planted dense subgraph \cite{AV13} and planted partition \cite{Bo87,CK01,FK01}.

There are two natural problems that arise in context of the SBM: \textit{exact
recovery}, where the aim is to recover the hidden partition completely; and
\textit{detection}, where the aim is to recover the partition better than what a
random guess would achieve. In this paper we focus on exact
recovery. Note that exact recovery necessarily requires the hidden clusters to
be connected (since otherwise there would be no way to match the partitions in one
component to another component) and it is easy to see that the threshold for connectivity occurs
when $p = \Omega\left(\log(m)/m\right)$. Therefore the right scale for
the threshold behavior of the parameters $p,q$ is
$\Theta\left(\log(m)/m\right)$, which is what we consider in this paper.

In the case of two communities ($k=2$) Abbe et al. \cite{ABH14} recently
established a sharp phase transition phenomenon from information-theoretic
impossibility to computational feasibility of exact recovery. However, the
existence of such a phenomenon in the case of $k>2$ was left open until solved,
for $k=O(1)$, in independent parallel research \cite{AS15,HWX15}. In this paper
we resolve the above showing the existence of a sharp phase transition for $k =
o(\log(n))$.

More precisely, in this work, we study a Semidefinite Programming (SDP) based algorithm that, for $k = o(\log(n))$, recovers, for an optimal range of parameters, exactly
the planted $k$-partition of $G \sim \mathcal{G}_{p,q,k}$ with high probability.
The range of the parameters $p,q$ is optimal in the following sense: it can be
shown that this parameter range exhibits a sharp phase transition from
information-theoretic impossibility to computational feasibility through the SDP
algorithm studied in this paper. An interesting aspect of our result is that,
for $k = o(\log(n))$, the threshold is the same as for $k=2$. This means that, even if an oracle reveals all of the cluster memberships except for two, the problem has essentially the same difficulty.
We also consider the case when $k = \Theta(\log(n))$. Unfortunately, in this regime we can no longer
guarantee exact recovery up to the proposed information theoretic threshold.
Similar behavior was observed and reported by Chen et al. \cite{CX14} and in our work
we observe that the divergence between our information theoretic
lower bound and our computational upper bound sets in at $k = \Theta(\log(n))$. This is formally summarized in the following theorems.

\begin{theorem}
\label{thm:main_theorem_1_opt}
Given a graph $G \sim \mathcal{G}_{p,q,k}$ with $k=O(\log(m))$ hidden
clusters each of size $m$ and
$p = \frac{\alpha \log(m)}{m}$ and $q = \frac{\beta \log(m)}{m}$, where $\alpha > \beta > 0$ are
fixed constants, the semidefinite program ~\eqref{eqn:sdp},
with probability $1 - n^{-\Omega(1)}$, recovers the clusters when: 
\begin{itemize}
\item for $k = o(\log n)$, as long as 
\[
\sqrt{\alpha} - \sqrt{\beta} > 1;
\]
\item for $k = \left( \gamma+o(1)\right) \log (n) $ for a fixed $\gamma$, as
long as
\[
\sqrt{\alpha} - \sqrt{\beta} > \sqrt{1 + c\sqrt{\beta\gamma} \left(1  +
\log\left(\sqrt{\frac{\alpha}{\beta}}\right)\right)},
\]
where $c$ is a universal constant.

\end{itemize}
\end{theorem}


We complement the above theorem by showing the following lower bound which is a
straightforward extension of the lower bound for $k=2$ from \cite{ABH14}. 

\begin{theorem}
\label{thm:main_theorem_1_lower_bound}
Given a graph $G \sim \mathcal{G}_{p,q,k}$ with $k$ hidden
clusters each of size $m$ where $k$ is $o(m^{-\lambda})$ for any fixed $\lambda
> 0$, if $p = \frac{\alpha \log(m)}{m}$ and $q = \frac{\beta \log(m)}{m}$, where
 $\alpha > \beta > 0$ are fixed constants, then it is information theoretically
 impossible to recover the clusters exactly with high probability if
 \[ \sqrt{\alpha} - \sqrt{\beta} < 1 \:.\]
\end{theorem}

Note that Theorem
\ref{thm:main_theorem_1_lower_bound} establishes a sharp phase transition between
computational feasibility and information theoretic impossibility when $k =
o(\log(n))$. At $k \sim \log(n)$ we see that our lower and upper bounds diverge.
We leave as an open problem to determine whether such divergence is necessary or
a shortcoming of the SDP approach.

At the heart of our
argument is the following theorem which establishes a sufficient condition for exact recovery with high probability.

\begin{theorem}
\label{thm:main_theorem_1}
Let $G \sim \mathcal{G}_{p,q,k}$, with probability $1 - n^{-\Omega(1)}$ over the
choice of $G$, if the following condition is satisfied, the semidefinite program
~\eqref{eqn:sdp} recovers the hidden partition:
\begin{equation}
\label{eqn:conditionmain}
\min_{i} \Delta(i) \geq \hat{c}\left(\sqrt{pn/k + qn} +
q\sqrt{\frac{n}{k}\log(n)} + \sqrt{\log(n)} + \log(k)\right) \:,
\end{equation}
where $\hat{c}$ is a universal constant and $\Delta(i)$ is defined as the
difference between the number of neighbors a vertex $i$ has in its own cluster
and the maximum number of neighbors it has in any other cluster (with respect
to the hidden partition).
In other words, with probability $1 - n^{-\Omega(1)}$, \eqref{eqn:conditionmain}
implies exact recovery.
 \end{theorem}

We are able to give sharp guarantees for the semidefinite programming algorithm
based essentially on the behavior of inner and outer degrees of the vertices.
This is achieved by constructing a candidate dual certificate and using bounds on the spectral norm of random matrices to show that the constructed candidate is indeed a valid one. The problem is then reduced to the easier task of understanding the typical values of such degrees. Remarkably, the conditions required for these quantities are very similar to the ones required for the problem to be information-theoretically solvable (which essentially correspond to each node having larger in-degree than out-degree). This helps explain the optimality of our algorithm. The approach of reducing the validity of a dual certificate to conditions on an interpretable quantity appeared in~\cite{Bandeira_Laplacian_2015} for a considerably simpler class of problems where the dual certificate construction is straightforward (which includes the stochastic block model for $k=2$ but not $k>2$). In contrast, in the current setting, the dual certificate construction is complex, rendering a different, and considerably more involved analysis. Moreover, the estimates we need (both of spectral norms and of inner and outer degrees) do not fall under the class of the ones studied in~\cite{Bandeira_Laplacian_2015}.

We also show that our
algorithm recovers the planted partitions exactly also in the presence of a
monotone adversary, a semi-random model defined in \cite{FK01}.

\subsection{Related Previous and Parallel Work}

Graph partitioning problem has been
studied over the years with various different objectives and guarantees. 
There has been significant recent literature concentration around the bipartiton
(bisection) and the general $k$-partition problems (multisection) in
random and semi-random models (\cite{DKMZ11}, \cite{MNS12},
\cite{MNS13}, \cite{YP14}, \cite{MNS14-2}, \cite{Ma14}, \cite{ABH14}, \cite{CX14}, \cite{MNS14}, \cite{Vu14}, \cite{CRV15}).

Some of the first results on partitioning random graphs were due to Bui et al. \cite{BCLS84} who presented
algorithms for finding bipartitions in dense graphs. Boppana \cite{Bo87} showed a spectral algorithm that for a
large range of parameters recovers a planted bipartition in a graph. Feige and
Kilian \cite{FK01} present an SDP based algorithm to solve the problem of
planted bipartition (along with the problems of finding Independent Sets and
Graph Coloring). Independently, McSherry \cite{Mc01} gave a spectral algorithm that solved the problems of Multisection, Clique and Graph Coloring.

More recently, a spate of results have established very interesting
phase transition phenomena for SBMs, both for the case of \textit{detection} and
\textit{exact} recovery. For the case of detection, where the aim is to recover
partitions better than a random guess asymptotically, recent works of
\cite{MNS12,MNS13,Ma14} established a striking sharp phase transition from
information theoretic impossibility to computational feasibility for the case of
$k=2$. For the case of exact recovery Abbe et al.
\cite{ABH14}, and independently \cite{MNS14}, established the existence of
a similar phase transition phenomenon albeit at a different parameter range.
More recently the same phenomenon was shown to exist for a semidefinite
programming relaxation, for $k=2$ in~\cite{HWX14,Bandeira_Laplacian_2015}.
However, the works described above established phase transition for $k=2$ and the
case for larger $k$ was left open. Our paper bridges the gap for larger
$k$ upto $o(\log(n))$ for the case of exact recovery. To put our work into
context, the corresponding case of establishing such behavior for the problem of
detection remains open. In fact, it is conjectured in \cite{DKMZ11,MNS12} that,
for the detection problem, there exists a gap between the thresholds for computational feasibility and
information theoretic impossibility for any $k$ number of communities greater than 4. In this
paper, we show that that is not case for the exact recovery problem.

Chen et al. \cite{CX14} also study the $k$-community SBM and
provide convex programming based algorithms and information theoretic lower
bounds for exact recovery. Their results are similar to ours in the sense that
they also conjecture a separation between information theoretic impossibility and
computation feasibility as $k$ grows. In comparison we focus strongly on the
case of slightly superconstant $k$ ($o(\log(n))$) and mildly growing $k$
($\Omega(\log(n))$) and show exact recovery to the optimal (even up to constants) threshold in the former case. Very
recently in independent and parallel work, Abbe and Sandon \cite{AS15} studied
the problem of exact recovery for a fixed number of ($k>2$) communities where the
symmetry constraint (equality of cluster sizes and the probabilities of connection are same in different clusters) is removed. 
Our result, in contrast to theirs, is based on the integrality of a semidefinite relaxation, which has the added benefit of producing an 
explicit certificate for optimality (i.e. indeed when the solution is ``integral'' we know for sure that it is the optimal balanced $k$-partition).
Abbe and Sandon \cite{AS15} comment in their paper that their results can be
extended for slightly superconstant $k$ but leave it as future work. In
another parallel and independent work, Hajek et al. \cite{HWX15} study
semidefinite programming relaxations for exact recovery in SBMs and achieve
similar results as ours. We remark that semidefinite program in consideration in
\cite{HWX15} is the same as the semidefinite program~\eqref{eqn:sdp} considered
by us (up to an additive/multiplicative shift) and both works
achieve the same optimality guarantee for $k=O(1)$. They also consider the problem of
SBM with 2 unequal sized clusters and the Binary Censored Block Model. In
contrast we show that the guarantees extend to the case even $k$ is
superconstant $o(\log(n))$ and provide sufficient guarantees for the case
of $k=\theta(\log(n))$ pointing to a possible divergence between information
theoretic possiblity and computational feasibility at $k=\log(n)$ which we leave
as an open question.
\subsection{Preliminaries}
\label{sec:preliminaries}
In this section we describe the notation and definitions which we use through
the rest of the paper. \\

\noindent\textbf{Notation.} Throughout the rest of the paper we will be
reserving capital letters such as $X$ for matrices and with $X[i,j]$ we will
denote the corresponding entries. In particular, $J$ will be used to denote the
all ones matrix and $I$ the identity matrix. Let $A \bullet B$ be the element
wise inner product of two matrices, i.e. $A \bullet B = Trace(A^TB)$. We note
that the all the logarithms used in this paper are natural logarithms i.e. with
the base $e$.

Let $G = (V,E)$ be a graph, $n$ the number of vertices and $A(G)$ its adjacency matrix. 
With $G \sim \mathcal{G}_{p,q,k}$ we denote a graph drawn from the stochastic
block model distribution as described earlier with $k$ denoting the
number of hidden {\em clusters}  each of size $m$. We denote
the underlying hidden partition with $\{P_t\}$. Let $P(i)$ be the function that
maps vertex $i$ to the cluster containing $i$. To avoid confusion in the
notation note that with $P_t$ we denote the $t^{th}$ cluster and $P(i)$ denotes
the cluster containing the vertex $i$.
We now describe the definitions of a few quantities which will be useful in
further discussion of our results as well as their proofs.
Define $\delta_{i \rightarrow P_t}$ to be the ``degree'' of vertex
$i$ to cluster $t$. Formally
\begin{equation*} \delta_{i \rightarrow P_t}\defeq \sum_{j \in P_t} A(G)[i,j]
\end{equation*} 
Similarly for any two clusters $P_{t_1},P_{t_2}$ define $\delta_{P_{t_1}
\rightarrow P_{t_2}}$ as 
\begin{equation*} \delta_{P_{t_1} \rightarrow P_{t_2}}\defeq \sum_{i \in
P_{t_1}}\sum_{j \in P_{t_2}} A(G)[i,j] \:. \end{equation*} 

Define the ``in degree'' of a vertex $i$, denoted
$\delta^{in}(i)$, to be the number of edges of going from the vertex to its own
cluster \begin{equation*} \delta^{in}(i) \defeq \delta_{i \rightarrow P(i)}
\:,\end{equation*} also define $\delta^{out}_{\max}(i)$ to be the maximum ``out
degree'' of a vertex $i$ to any other cluster
\begin{equation*} \delta^{out}_{\max}(i) \defeq \max\limits_{P_t \neq P(i)}
\delta_{i \rightarrow P_t} \:.\end{equation*}
Finally, define \[\Delta(i) \defeq \delta^{in}(i) - \delta_{\max}^{out}(i)\:, \]
$\Delta(i)$ will be the crucial parameter in our threshold.
Remember that $\Delta(i)$ for $A(G)$ is a random variable and let $\Delta \defeq
\expect[\Delta(i)]$ be its expectation (same for all $i$). \\

\noindent\textbf{Paper Organization.} The rest of this paper is structured as
follows. In Section \ref{sec:sdps} we discuss the two SDP relaxations we
consider in the paper. We state sufficient conditions for exact recovery for
both of them as Theorem \ref{thm:main_theorem_sdp2} and Theorem
\ref{thm:main_theorem_1_restate} (the latter is a restatement of Theorem
\ref{thm:main_theorem_1}) and provide an intuitive explanation of why the
condition \eqref{eqn:conditionmain} is sufficient for recovery upto the optimal
threshold. We provide formal proofs of Theorems \ref{thm:main_theorem_1_opt} and
\ref{thm:main_theorem_1_lower_bound} in the Appendix in Sections
\ref{sec:proofmaintheoremupperbound} and \ref{sec:proofmaintheoremlowerbound}
respectively.
We provide the proof of Theorem \ref{thm:main_theorem_1_restate} in Section
\ref{sec:proof_main_theorem}.
Further in Section \ref{sec:adversary} we show how our result can be
extended to a semi random model with a monotone adversary. Lastly in the
Appendix we collect the proofs of all the lemmas and theorems left unproven in
the main sections.

\section{SDP relaxations and main results}
\label{sec:sdps}

In this section we present two candidate SDPs which we use to recover the hidden
partition. The first SDP is inspired from the Max-k-Cut SDP introduced by Frieze
and Jerrum \cite{FJ95} where we do not explicitly encode the fact that each
cluster contains equal number of vertices. In the second SDP we encode the fact
that each cluster has exactly $m$ vertices explicitly. We state our main
theorems which provide sufficient conditions for exact recovery in both SDPs.
Indeed the latter SDP, being stronger, is the one we use to prove our main theorem, Theorem~\ref{thm:main_theorem_1_opt}.
Before describing the SDPs lets first consider the Maximum Likelihood Estimator
(MLE) of the hidden partition.
It is easy to see that the MLE corresponds to the following problem which we refer
to as the Multisection problem. Given a graph $G=(V,E)$ divide the set of
vertices into $k$ clusters $\{P_t\}$ such that for all
$t_1,t_2$, $|P_{t_1}|=|P_{t_2}|$ and the number of edges $(u,v) \in E$ such that
$u \in P_{t_1}$ and $v \in P_{t_2}$ are minimized. (This problem has been
studied under the name of Min-Balanced-k-partition \cite{KNS09}). In this
section we consider two SDP relaxations for the Multisection problem. Since
SDPs can be solved in polynomial time, the 
relaxations provide polynomial time algorithms to recover the hidden partitions. 
 
A natural relaxation to consider for the problem of multisection in the
Stochastic Block Model is the Min-k-cut SDP relaxation studied by
Frieze and Jerrum \cite{FJ95} (They actually study the Max-k-Cut problem but we
can analogously study the min cut version too). The Min-k-cut SDP formulates the
problem as an instance of Min-k-cut where one tries to separate the graph into $k$ partitions 
with the objective of minimizing the number of edges cut by the partition. Note
that the k-Cut version does not have any explicit constraints for ensuring balancedness.
However studying Min-k-Cut through SDPs has a natural difficulty, 
the relaxation must explicitly contain a constraint that tells it to divide the
graph into at least $k$ clusters. In the case of SBMs with the parameters
$\alpha\frac{\log(n)}{n}$ and $\beta\frac{\log(n)}{n}$ one can try and overcome
the above difficulty by making use of the fact that the generated graph is
very sparse. Thus, instead of looking directly at the min-k-cut objective we can
consider the following objective: minimizing the difference between the number
of edges cut and the number of non-edges cut. Indeed for sparse graphs the
second term in the difference is the dominant term and hence the SDP has an incentive to produce more clusters. Note that the above objective
can also be thought of as doing Min-k-Cut on the signed adjacency matrix $2A(G)
- J$ (where $J$ is the all ones matrix). Following the above intuition we
consider the following SDP \eqref{eqn:sdp2} which is inspired from the Max-k-Cut
formulation of Feige and Jerrum \cite{FJ95}. In the Appendix Section
\ref{sec:nonuniquegames} we provide a reduction, to the k-Cut SDP we study in this paper, from a more general class of SDPs studied by Charikar et
al. \cite{CMM06} for Unique Games, and more recently by Bandeira et al.
\cite{BCS15} in a more general setting. 
 
\begin{equation} \label{eqn:sdp2} \begin{tabular}{l p{4cm} l}
    $\max$ & ${\displaystyle (2A(G) - J)\bullet Y }$ & \\ 
    s.t.  & ${\displaystyle Y_{ii} = 1\;\; (\forall\;i)}$ \\
    & ${\displaystyle Y_{ij} \geq -\frac{1}{k-1}\;\; (\forall\;i,j)}$ \\
    & ${\displaystyle Y \succcurlyeq 0 }$ \:.
\end{tabular} \end{equation}

To see that the above SDP is a relaxation of the multisection problem note that
for the hidden partition $\{P_t\}$ we can define a candidate solution $Y^*$ as
follows. $Y^*_{ij} = 1$ if $i,j$ belong
to the same cluster and $-\frac{1}{k-1}$ if $i,j$ belong to different
clusters. Note that although the objective does not directly minimize the
number of edges cut, it is an additive/multiplicative shift of it. For the
above SDP we prove the following theorem in the Appendix in Section
\ref{sec:proofSDP2}.
Given $G \sim \mathcal{G}_{p,q,k}$, define

\[ \nu(i) \defeq \delta_{in}(i) - \max_{i,j}\left(\delta_{i \rightarrow P(j)} +
\delta_{j \rightarrow P(i)} - \frac{\delta_{P(j) \rightarrow P(i)}}{n/k}\right)
\]

\begin{theorem}
\label{thm:main_theorem_sdp2}
Let $G \sim \mathcal{G}_{p,q,k}$, with $p = \alpha\frac{\log(m)}{m}$ and $q =
\beta\frac{\log(m)}{m}$ where $\alpha,\beta$ are constant. Consider the SDP
given by \eqref{eqn:sdp2}.
With probability $1 - n^{-\Omega(1)}$ over the choice of $G$, if the following condition is satisfied then the SDP recovers the hidden partition
\begin{equation}
\label{eqn:sdp2_condition}
\min_{i} \nu(i) \geq \hat{c}\left(\sqrt{pn/k + qn} + \sqrt{\log(n)}\right) \:,
\end{equation}
where $\hat{c}$ is a universal constant.

In other words with probability $1 - n^{-\Omega(1)}$, condition~\eqref{eqn:sdp2_condition} implies exact recovery.
\end{theorem}
 
The proof of the above Theorem is included in the Appendix in Section
\ref{sec:proofSDP2}. We note the above condition is not an optimal one in terms
of exact recovery and we discuss this issue next. It is quite possible that the above
 SDP recovers the planted multisection all the way down to the threshold however we have not been
able to establish this and leave it as an open question.
Indeed to prove our results we consider a stronger SDP with which we establish
optimality. We have empirically tested the performance of both the SDPs and
include the results in the Appendix in Section \ref{sec:experiments}. We now take a closer look at
the above sufficient condition \eqref{eqn:sdp2_condition} and argue
why the condition is not strong enough to achieve optimal results.
It is not hard to see that 
\[ \expect[\nu(i)] \sim p\frac{n}{k} - q\frac{n}{k} -
O\left(\sqrt{q\frac{n}{k}\log(n)}\right) \] 
Note that, in expectation, the
maximization term in the definition of $\nu(i)$ has an extra $\log(n)$ term as the maximization runs through all $i,j$ pairs. For the
condition \eqref{eqn:sdp2_condition} to hold with at least a constant
probability, we expect that it needs to be the case that 
\[p\frac{n}{k} - q\frac{n}{k} - O\left(\sqrt{q\frac{n}{k}\log(n)}\right) \geq
O\left(\sqrt{p\frac{n}{k} + q\frac{n}{k}k} + \sqrt{\log(n)}\right)\]
Substituting the parameter range that we are interested $p =
\frac{\alpha\log(m)}{m}$ and $q = \frac{\beta\log(m)}{m}$ we require that 
\[ \alpha - \beta \geq O\left(\sqrt{\beta} + \sqrt{\frac{\beta
k}{\log(n)}}\right)\] Indeed from the above expression it is clear that if $k <<
\log(n)$ the first term above dominates and we cannot expect to get the tight results we hope for
in Theorem \ref{thm:main_theorem_1_opt}. A closer look at the above calculation
reveals that the major barrier towards achieving the optimal result is the
additional $\log(n)$ factor due to the maximization over all $i,j$ in
the definition of $\nu(i)$. For instance if one could replace the maximization
term above with a term that takes the maximum per vertex over all clusters one
would pick up only a $\log(k)$ term (as there are only $k$ clusters) and
hopefully achieve optimality.

In context of the above discussion we suggest the following
SDP in which we explicitly add a per-row constraint bounding the number
of vertices belonging to the same cluster as the vertex in contention.

\begin{equation} \label{eqn:sdp} \begin{tabular}{l p{4cm} l}
    $\max$ & ${\displaystyle A(G)\bullet Y }$ & \\ 
    s.t. & ${\displaystyle \sum_{j} Y_{ij} + \sum_{j} Y_{ji}  = 2n/k  \;\;
    (\forall\;i)}$ \\
    & ${\displaystyle Y_{ii} = 1\;\; (\forall\;i)}$ \\
    & ${\displaystyle Y_{ij} \geq 0\;\; (\forall\;i,j)}$ \\
    & ${\displaystyle Y \succcurlyeq 0 }$ \:.
\end{tabular} \end{equation}

To see that the above SDP is a relaxation of the MLE discussed above note that
for any partition $P = \{P_i\}$, we can associate a canonical
$n \times n$ matrix $Y_P$ with it defined as 
\begin{equation*}
    Y_P[i,j] = \left\{ \begin{array}{rl}
               1 & \mbox{ vertex $i$ and $j$ belong to the same cluster}\\
               0 & \mbox{ otherwise}
               \end{array} \right.
\end{equation*} Note that $Y_P$ satisfies the SDP constraints and the SDP
maximizes the number of edges within the cluster which is equivalent to
minimizing the number of edges across the clusters. The second constraint
above, since $Y$ is symmetric, says that the sum of the values along the row is $n/k$, 
which represents the number of vertices in a cluster.
For the SDP above we show the following theorem which is a restatement of
Theorem \ref{thm:main_theorem_1}

\begin{theorem}
\label{thm:main_theorem_1_restate}
Let $G \sim \mathcal{G}_{p,q,k}$. With probability $1 - n^{-\Omega(1)}$ over the
choice of $G$, if the following condition is satisfied then the SDP defined by
\eqref{eqn:sdp} recovers the hidden partition
\begin{equation}
\label{eqn:condition}
\min_{i} \Delta(i) \geq \hat{c}\left(\sqrt{pn/k + qn} +
q\sqrt{\frac{n}{k}\log(n)} + \sqrt{\log(n)} + \log(k)\right) \:,
\end{equation}
In other words with probability $1 - n^{-\Omega(1)}$, condition~\eqref{eqn:condition}
implies exact recovery.
\end{theorem}

We remark that the above statement is indeed true for all values of $p,q$. For
the specific range that we are interested in we show in Section~\ref{sec:proofmaintheoremupperbound} how condition~\eqref{eqn:condition}
leads to the optimal threshold. In the next section we provide an intuitive
explanation of why this is so.

\subsection{Optimality of Theorem \ref{thm:main_theorem_1_restate}}
\label{sec:optimality}

In this section we give an intuitive high level explanation for the optimality
of the condition in \eqref{eqn:condition} for $k << \log(n)$ in
Theorem \ref{thm:main_theorem_1_restate}. We prove it formally in the appendix.
As stated earlier the regime we consider is the case when $p = \frac{\alpha \log(m)}{m}$ and
$q = \frac{\beta \log(m)}{m}$, where $\alpha$ and $\beta$ are
constants. 

Note that for the MLE to succeed the values of $p$ and $q$ should be such that
$\min_{i}\{\delta^{in}(i) - \delta_{\max}^{out}(i)\} \geq 0$ w.h.p., since otherwise one expects there to be many such vertices $i$ for which
$\delta^{in}(i) - \delta_{i \rightarrow P_t} \leq 0$ for some $t \neq P(i)$ and 
in particular a pair $t_1,t_2$ such that there exists $i \in P_{t_1},j \in P_{t_2}$ such that $\delta^{in}(i) - \delta_{i \rightarrow P_{t_2}} \leq 0$ as well as 
$\delta^{in}(j) - \delta_{i \rightarrow P_{t_1}} \leq 0$. This would imply that we can exchange
the pairs $i,j$ and get a better partition than the planted partition and
therefore that the MLE itself does not recover the hidden partition.

Recall that $\Delta(i)=\delta^{in} -
\delta_{\max}^{out}(i)$. We now show that the deviation in $\Delta(i)$ required
by Theorem \ref{thm:main_theorem_1_restate} is
$o\left(\mathbb{E}[\Delta(i)]\right)$ and therefore informally one can expect, intuitively, 
that \[\mathbb{P}(\min_i \Delta(i) \geq 0) \sim \mathbb{P}\left(\min_i \Delta(i) \geq
o\left(\mathbb{E}[\Delta(i)]\right) \right)\] 
which implies that the SDP in Theorem \ref{thm:main_theorem_1_restate} recovers
the partition optimally. Indeed, the
deviation required in Theorem \ref{thm:main_theorem_1_restate} is
$o\left(\mathbb{E}[\Delta(i)]\right)$,

\begin{align*}
\frac{\left(\sqrt{pn/k + qn} +
q\sqrt{n/k\log(n)} + \sqrt{\log(n)}\right)}{\mathbb{E}[\Delta(i)] } &= \frac{O\left(
\sqrt{\log(m)(\alpha + k\beta)}\right) + O(\sqrt{\log(n)})}{\Omega\left((\alpha
- \beta)\log(m)\right) }\\
 &= o(1) \:.
\end{align*}

Above we assumed that $k = o(\log(n))$. Following from the intuition above we prove Theorems
\ref{thm:main_theorem_1_opt} and \ref{thm:main_theorem_1_lower_bound} in the
appendix which imply that our SDP is optimal.

In the Appendix (Section \ref{sec:experiments}) we present an experimental
evaluation of the two SDPs considered in this section. The experiments
corroborate Theorem \ref{thm:main_theorem_1_opt} and also
show that the SDP in \eqref{eqn:sdp2} experimentally seems to have a similar
recovery performance as the (stronger) SDP in \eqref{eqn:sdp} however we could
only prove a suboptimal result about it.
We leave the possible optimality of the SDP in \eqref{eqn:sdp2} as an open question.

\section{Proof of the main theorem}

\label{sec:proof_main_theorem}

In this section we prove our main theorem, Theorem~\ref{thm:main_theorem_1_restate} about
the SDP defined by \eqref{eqn:sdp}. We restate the SDP here. 

\begin{equation} \begin{tabular}{l p{4cm} l}
    $\max$ & ${\displaystyle A(G)\bullet Y }$ & \\ 
    s.t. & ${\displaystyle \sum_{j} Y_{ij} + \sum_{j} Y_{ji}  = 2n/k  \;\;
    (\forall\;i)}$ \\
    & ${\displaystyle Y_{ii} = 1\;\; (\forall\;i)}$ \\
    & ${\displaystyle Y_{ij} \geq 0\;\; (\forall\;i,j)}$ \\
    & ${\displaystyle Y \succcurlyeq 0 }$ \:.
\end{tabular} \end{equation}

  Let $Y^*$ be the matrix corresponding to the hidden partition $P^* =
  \{P_t\}$, i.e.
  $Y^*[i,j] = 1$ if $i,j$ belong to the same cluster and $0$ otherwise. Let
  $OPT(G)$ be the optimal value in the above SDP.
We will show that $Y^*$ is the unique solution to SDP (\ref{eqn:sdp}) w.h.p as
long as the conditions in Theorem \ref{thm:main_theorem_1_restate} are satisfied. This would prove Theorem
\ref{thm:main_theorem_1_restate}. Our proof will be based on a dual
certificate. In that context consider the dual formulation of the above SDP
which is the following

\begin{equation}\label{eqn:dual} \begin{tabular}{l p{4cm} l}
    min & ${\displaystyle \mbox{Trace}(D) + (2n/k)\sum_i x_i }$ & \\ 
    s.t. & ${\displaystyle D + \sum_i x_i (R_i + C_i) - Z - A\succcurlyeq 0 \:. }$
\end{tabular} \end{equation} 
where $D$ is a diagonal matrix, $x_i$ are scalars, $Z$ is a non-negative
symmetric matrix (corresponding to the $\geq 0$ constraints) with $0$
in the diagonal entries, $R_i$ is the matrix with $1$ in every entry of row $i$ and
0 otherwise, $C_i=R_i^T$ is the matrix with $1$ in every entry of column $i$ and
0 otherwise and we write $A$ instead of $A(G)$ when there is no fear of confusion.

Let $DUAL(G)$ be the optimal value of the above dual program. We will first exhibit a valid dual solution
$M^* = (D^*,\{x_i^*\},Z^*)$ which, with high
probability, has dual objective value $\delta$ such that $A \bullet
Y^* = \delta$. But since $A \bullet Y^* \leq OPT(G) \leq DUAL(G)$ (by weak
duality) we get that $Y^*$ is an optimal solution to the above SDP. We will also show uniqueness via complementary slackness. 

Before moving on further it will be convenient to introduce the following
definition which we will be used in the proof later. We also encourage the
reader to revisit the Notations section (Section~\ref{sec:preliminaries}) at this time as it would
help with the reading of what follows.


\begin{definition}
\label{def:spaces}
Given a partition of $n$ vertices $\{P_t\}_{t=1}^k$ we define the vectors
$\{v_t\}$ to be the indicator vectors of the clusters. We further define the
following subspaces, which are perpendicular to each other, and partition $\R^n$.
\begin{itemize}
\item $\R_{k}$: the subspace spanned by the vectors $\{v_t\}$, i.e. the
subspace of vectors with equal values in each cluster,
\item $\R_{n | k}$: the subspace perpendicular to $\R_k$, i.e. the subspace
where the sum on each cluster is equal to 0.
\end{itemize}
\end{definition}


At this point it is useful to look at what the complementary slackness condition
implies. Since strong duality holds in the case of our SDP (easy to check that
Slater's conditions are satisfied) we have that complementary slackness is zero
which implies that \[\mbox{Trace}(M^*Y^*) = \mbox{Trace}\left(M^*\sum
v_tv_t^T\right) = 0 \:.\] for any optimal dual solution $M^*$. The above
condition implies that for any such $M^*$ (since $M^*$ is PSD) it
must be that the subspace $\R_k$ is an eigenspace with eigenvalue $0$ which implies 
\begin{equation}
\label{eqn:sumcondition}
(\forall i,t) \delta_{i \rightarrow P_t}(M^*) = 0\:.
\end{equation}

Having established the conditions that must be satisfied by the optimal dual
solution $M^*$, we describe our candidate dual solution \[ (D^*, \{ x_i^*\}, Z^*)\:. \]

We begin by describing the choice of $Z^*$. If vertex $i$ and $j$ belong to the
same cluster then $Z^*[i,j] = 0$ otherwise
\begin{equation*}
    Z^*[i,j] = \left( \frac{\delta^{out}_{\max}(i)}{n/k} - \frac{\delta_{i
               \rightarrow P(j)}}{n/k} \right)+
               \left( \frac{\delta^{out}_{\max}(j)}{n/k} - \frac{\delta_{j
               \rightarrow P(i)}}{n/k} \right) +
               \left(\frac{\delta_{P(j) \rightarrow P(i)}}{(n/k)(n/k)} -
               \min_{t_1,t_2} \frac{\delta_{P_{t_1} \rightarrow
               P_{t_2}}}{(n/k)(n/k)} \right) \:.               
\end{equation*}

It is easy to see that the matrix $Z^*$ is symmetric by noting that exchanging
$j$ and $i$ in the above expression leads to the same value. Also to see that each
entry of $Z^*$ is non-negative note that $Z^*[i,j]$ is the sum of non-negative terms. Having defined $Z^*$ as above we choose $x_i^*$ to be such that
the condition given in Equation \ref{eqn:sumcondition} holds for the non-diagonal blocks, yielding:

\begin{equation*}
x_i^* = \frac{\delta^{out}_{\max}(i)}{n/k} - \frac{1}{2} \min_{t_1,t_2}
\frac{\delta_{P_{t_1} \rightarrow P_{t_2}}}{(n/k)(n/k)}\:.
\end{equation*}

And finally we define $D^*$ to balance out the sum along the diagonal blocks
from $A$ as well as the $x^*_i$.
\begin{equation*}
D^*[i,i] = \delta^{in}(i) - \delta^{out}_{\max}(i) -
\sum_{j \in P(i)} \frac{\delta^{out}_{\max}(j)}{n/k} + \min_{t_1,t_2}
\frac{\delta_{P_{t_1} \rightarrow P_{t_2}}}{n/k}\:.
\end{equation*}

Interestingly, this dual certificate construction seems to share some features with the one proposed by Awasthi et al.~\cite{Awasthi_KmeansSDP} for an SDP relaxation for k-means clustering. While we were not able to make a formal connection, it would be very interesting if the reason for the similarities was the existence of some type of canonical way of building certificates for clustering problems, we leave this for future investigations.

Now consider the objective for the dual
program \eqref{eqn:dual}. It is easy to see that it is equal to
\begin{equation*} \mbox{Trace}(D^*) + 2n/k\sum_i x^*_i = \sum_i \delta^{in}(i) = A(G) \bullet Y^*
\:. \end{equation*} The following lemma the proof of which we provide in
the Appendix in Section \ref{sec:proof_main_lemma} implies that the above
mentioned solution is a valid dual solution, proving that $Y^*$ is an optimal solution to the above program (by weak duality).

\begin{lemma}
\label{lem:main_lemma}
The matrix $M^* = D^* + \sum_{i}x_i^*(R_i+C_i) - A - Z^*$ (as defined above)
is such that with probability $1 - n^{-\Omega(1)}$, if the condition
\eqref{eqn:condition} is satisfied, then 
\begin{equation*}M^* \succeq 0 \:. \end{equation*}
\end{lemma}

%
%

It is easy to show using complementary slackness that $Y^*$ is indeed the unique
optimal solution with high probability. For completeness we provide the proof
in the Appendix in Section \ref{sec:uniqueness}

\section{Note about the Monotone Adversary}
\label{sec:adversary}

In this section, we extend our
result to the following semi random model considered in the paper of Feige and Kilian \cite{FK01}. We first define a monotone
adversary (we define it for the ``homophilic'' case). Given a graph $G$ and a
partition $P = \{P_i\}$ a \textit{monotone} adversary is allowed to
take any of the following two actions on the graph:
\begin{itemize}
  \item Arbitrarily remove edges across clusters, i.e. $(u,v) \mbox{ s.t. }
  P(u) \neq P(v)$.
  \item Arbitrarily add edges within clusters, i.e. $(u,v) \mbox{ s.t. } P(u)
  = P(v)$.
\end{itemize} 
Given a graph $G$ let $G_{adv}$ be the resulting graph after the adversary's
actions. The adversary is monotone in the sense that the set of the
optimal multisections in $G_{adv}$ contains the set of the
optimal multisections in $G$. Let $B(G)$ be the number of edges cut in
the optimal multisection. We now
consider the following semi-random model, where we first randomly pick a graph
$G \sim \mathcal{G}_{p,q,k}$ and then the algorithm is given $G_{adv}$ where the
monotone adversary has acted on $G$. The following theorem shows that our algorithm is robust against such a monotone adversary:
\begin{theorem}
\label{thm:adversary}
Given a graph $G_{adv}$ generated by a semi-random model described above we have
that with probability $1-o(1)$ the algorithm described in section
\ref{sec:proof_main_theorem} recovers the original (hidden) partition. The probability
is over the randomness in the production of $G \sim \mathcal{G}_{p,q,k}$ on
which the adverary acts. 
\end{theorem}

We provide the proof of the above theorem in the Appendix in Section
\ref{sec:adversaryproof}

\bibliographystyle{alpha}
\bibliography{partitioning}
\ENDDOC